# GaSb-based Interband Cascade Laser with hybrid superlattice plasmon-enhanced claddings

B. Petrović,[1,a] A. Bader,[1] J. Nauschütz,[2] T. Sato,[3] S. Birner,[3] R. Weih,[2] F. Hartmann,[1,a] and S. Höfling[1]

[1]Julius-Maximilians-Universität Würzburg, Physikalisches Institut, Lehrstuhl für Technische Physik, Am Hubland, 97074 Würzburg, Germany

[2]nanoplus Advanced Photonics Gerbrunn GmbH, Oberer Kirschberg 4, 97218 Gerbrunn, Germany

[3]nextnano GmbH, Konrad-Zuse-Platz 8, 81829 München, Germany

[a]Authors to whom correspondence should be addressed: borislav.petrovic@uni-wuerzburg.de and fabian.hartmann@uni-wuerzburg.de

We present an interband cascade laser (ICL) emitting at 5.2 µm consisting of an 8-stage active region and a hybrid cladding composed of outer plasmon-enhanced $InAs_{0.915}Sb_{0.085}$ and inner InAs/AlSb superlattice claddings. The hybrid cladding architecture shows an increase in mode-confinement in the active region by 11.2 % according to the simulation. This is a consequence of a significantly lower refractive index of plasmon-enhanced claddings. The threshold current density is 242 A/cm² in pulsed operation at room temperature. This is the lowest value reported to date for ICLs emitting at wavelengths longer than 5 µm. We also report close to record value threshold power density of 840 W/cm² for ICLs at such wavelengths.

Interband cascade lasers (ICLs[1]) are important coherent mid-infrared light sources for a variety of applications such as industrial process control, medical diagnosis, IR scene projection and defense[2-8]. Multiple photon generation for one injected electron-hole pair inherited from its cascading scheme provides low threshold current densities and high voltage efficiencies what makes ICLs attractive devices for portable use[8-9]. Especially in the 3-6 µm wavelength range, ICLs show lower threshold current and power densities then diode lasers or quantum cascade lasers[8-9] for example. The first continuous wave (cw) operation at room temperature and first pulsed operation with threshold current densities as low as 400 A/cm² demonstrated at a wavelength of 3.75 µm in 2008 was a milestone in ICL design[10]. Use of lightly doped separate confinement layers (SCLs) and moderate doping of the claddings was introduced and set as the state-of-the-art since then[8-9]. A significant reduction of threshold current density of broad area (BA) devices at room temperature (134 A/cm²) has been achieved at the "sweet spot" wavelength range of 3.6-3.7 µm by heavily doping the electron injectors to balance the excess population of holes in the recombination region[11]. Additional improvements were achieved by optimizing the number of cascade stages[12]. Moving towards longer wavelengths, reducing threshold current density becomes more challenging as the wavefunction overlap between the laser transiton levels decreases and free carrier absorption losses increase with wavelength squared according to Drude model.

In the last decade, a remarkable effort on ICLs emitting above 4 µm has been made. As a spectral line of nitrous-oxide (NO), the byproduct of fossil fuels, shows absorption at 5.2 µm, this wavelength is of particular interest for detecting air pollution. In 2014, the first cw operation of a distributed feedback (DFB) laser at 5.2 µm was demonstrated[13]. Previously reported pulsed threshold current densities at such wavelengths and at room temperature for 10 stage[14] and 5 stage[15] ICLs were 1.3 kA/cm² and 650 A/cm² respectively. There was a major improvement reported in 2015[16-17] for 5 stage ICLs and the threshold current density was decreased to 280 A/cm².

Over recent years, a prerequisite for emission at longer wavelengths (even up to 14.4 µm[18] at temperature of 120 K) has been the strong optical confinement in the active region. For this reason, ICLs grown on InAs substrates with highly doped plasmon enhanced InAs claddings have been used. In this paper, we are applying a similar concept for a 5.2 µm ICL grown on a GaSb substrate. The cladding structure design is analog to InAs-based ICLs reported for 4.6 µm emission[19], however n-type doped $InAs_{0.915}Sb_{0.085}$ plasmon-enhanced outer claddings were implemented instead of InAs, due to the possibility of lattice matching to the GaSb substrate. The inner cladding layers were composed of InAs/AlSb superlattices and SCLs were GaSb layers, which promise to confine light more effectively than InAs due to the higher refractive index contrast[20]. In this paper we report an additional reduction in threshold current density of ICL to 242 A/cm² at room temperature employing this design. It is noteworthy that previous values[16-17] were reported for operation at 20 °C, what makes the improvement of the threshold current density larger than 13 %.

An improvement in the optical confinement of highly doped ($1 \cdot 10^{19} cm^{-3}$) n⁺-InAsSb compared to InAs/AlSb superlattices is achieved by the reduced refractive index[21] of $n_{InAsSb}$ = 2.70 compared to $n_{SL}$ = 3.39 for InAs/AlSb superlattices[22]. To verify the benefits of such hybrid claddings, we compare the optical confinement factors calculated by Helmholtz wave equation solver of two analog ICL structures with different claddings. Figure 1 (a) displays the profiles of refractive indices ($n_r$)[20-23] and relative mode intensities (R.I.) of the two ICLs with the only difference being the cladding structure. Figure 1 (b) shows the optical mode profile with distribution of confinement factors for the ICL with hybrid claddings. The hybrid cladding thickness of 1.7 µm corresponds to the same order of magnitude of relative mode intensity in vicinity of the substrate as for the ICL with standard 3.3 µm and 2 µm thick InAs/AlSb superlattice claddings. Nevertheless, the ICL with hybrid claddings shows stronger confinement in the active region due to significantly lower refractive index of n$^+$-InAsSb layers. This reflects a 11.2 % increase in comparison to ICL with only superlattice claddings which is the main motivation for using a hybrid cladding design. However, based on doping levels, effective masses[24], mobilities[21,25-26] and optical confinement factors, the calculated overall free carrier absorption in claddings and SCLs of the hybrid architecture ($\alpha$ = 5.9 cm$^{-1}$) is higher than in the standard design ($\alpha$ = 4.5 cm$^{-1}$). This is a consequence of a larger mode confinement in SCLs and higher optical losses in n$^+$-InAsSb layers.

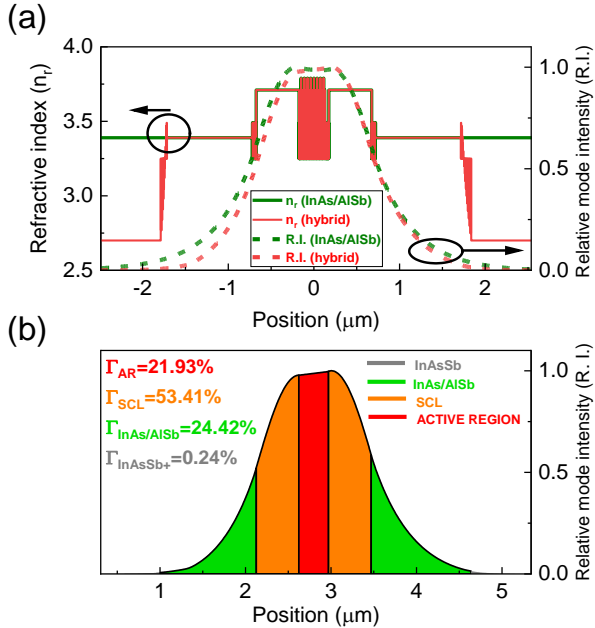

FIG. 1. (a) Refractive index profiles and optical mode intensities along the growth direction of two ICLs with the same SCLs and active region design, but different cladding designs: standard InAs/AlSb superlattice claddings (green) and hybrid InAs/AlSb - InAsSb plasmon-enhanced claddings (red) (b) Optical mode confinement distribution along the growth direction of the ICL with hybrid cladding designs.

The lower refractive index of n$^+$-InAsSb allows to reduce the thickness of the claddings. Additionally, substituting a part of the InAs/AlSb superlattice of low thermal conductivity (≈3 W/m·K)[27] with bulk n$^+$-InAsSb of a higher thermal conductivity (≈15 W/m·K)[27] in a hybrid-cladding-based ICL is likely to improve heat dissipation. The active region is designed for a laser emission wavelength of 5.2 µm and consists of InAs/Ga$_{0.6}$In$_{0.4}$Sb/InAs W-shaped quantum wells. In a single W-shaped quantum well, InAs wells confine electrons and sandwich the hole-confining GaInSb-well. In Ref. 28 a significant performance improvement was reported for a 4.35 µm GaSb-based ICL design by mitigating intervalence band absorption in the active region. A threshold current density of 247 A/cm$^2$ has been obtained in the range 4.3-4.6 µm. The study has shown that in the 4.0-4.8 µm range, optimal thickness of GaInSb hole-quantum well is 2.5 nm, whereas for above 5.4 µm it is 3.5 nm. This concept was later experimentally confirmed for a 6.2 µm design[29]. For our design we selected a GaInSb thickness of 3.0 nm for the target emission range 4.8-5.4 µm[28-29]. The InAs electron quantum well thickness has been determined by sweeping the widths of both electron and hole wells using the software nextnano[30]. The simulation is based on **k·p** band structure calculations using the finite difference method. The band parameters are obtained from Ref. 24. Figure 2 (a) shows the lasing wavelength versus GaInSb hole-quantum well and InAs electron-quantum well thickness variations. In Figure 2 (b) the corresponding wavefunction overlap is depicted for the same parameter range. For increasing thicknesses of InAs wells, the emission wavelength increases, whereas the wavefunction overlap decreases. The variation of the transition energy mostly depends on the widths of InAs electron quantum wells. The calculation further provides an almost constant overlap integral for varying the thickness of the GaInSb well assuring that there is no sacrifice from pre-selecting its value.

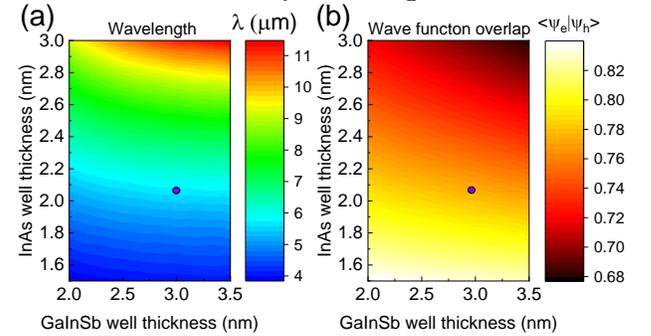

FIG. 2. Dependence of the emission wavelength (in (a)) and the corresponding wavefunction overlap (in (b)) for a variation of the electron (InAs) and hole (Ga$_{0.6}$In$_{0.4}$Sb) quantum well widths. For a fixed wavelength the wavefunction overlap remains constant and almost independent of the hole quantum well. The chosen design is labeled by the purple dot.

The ICL was grown by molecular beam epitaxy equipped with standard effusion cells for the group III elements, silicon, and tellurium as n-type dopants and two valved cracker cells for arsenic and antimony. The growth was carried out on a 2″ GaSb substrate n-type doped with Te ($1 \cdot 10^{18}$cm$^{-3}$). The substrate temperature during the growth was monitored by a pyrometer. An oxide desorption step was carried out under Sb flux at 560 ºC for 3 minutes. The growth began at a substrate temperature of 500 ºC with 200 nm thick n-type doped GaSb:Te buffer layer ($1 \cdot 10^{18}$ cm$^{-3}$). In Figure 3 (a) the schematic layout of the ICL is shown. The layer sequence starts with plasmon enhanced n$^+$-InAsSb:Si layer grown at lower temperature of 450 ºC which was maintained during the growth of the inner bottom InAs/AlSb cladding. The substrate temperature was then increased to 500 ºC for the growth of the bottom GaSb-SCL, then ramped down again to 450 ºC for the growth of the 8-stage active region and remains unchanged at 450 ºC for the rest of the growth. Outer claddings were 700 nm n$^+$-InAsSb layers ($1 \cdot 10^{19}$ cm$^{-3}$) and 1 µm long inner claddings were grown as a strain compensated superlattice with 212 periods of 2.43 nm/2.30 nm thick InAs/AlSb layers. The InAs layers of the SL claddings were Si-doped with a linearly decreasing doping concentration towards the active region ($1 \cdot 10^{18}$ cm$^{-3}$ to $1 \cdot 10^{17}$ cm$^{-3}$) to minimize optical losses. Both SCLs were 500 nm thick to increase the confinement in the active region and lightly Te-doped ($6 \cdot 10^{16}$ cm$^{-3}$) also for the sake of lower optical losses. The active W-QW is composed of **AlSb**/*InAs*/GaInSb/*InAs*/**AlSb** with thicknesses of (**2.5**/*2.12*/3.00/*1.98*/**1.0**) nm. The hole and electron injector are composed of 2.8 nm GaSb/1.0 nm AlSb/4.8 nm GaSb and 2.5 nm AlSb/4.4 nm InAs/1.2 nm AlSb/3.2 nm InAs/1.2 nm AlSb/2.5 nm InAs/1.2 nm AlSb/2.05 nm InAs respectively. For rebalancing the hole density in the active region[31], four InAs wells of the electron injectors were n-type doped to $4 \cdot 10^{18}$ cm$^{-3}$. The number of cascade stages was chosen by the values for plasmon-enhanced cladding based ICLs that have shown a good performance based on voltage efficiency[32]. Adding more cascade stages would increase the useful voltage and by reducing the threshold current it would decrease ohmic parasitic voltages. Consequently, the voltage efficiency would improve, however increasing overall voltage would also increase the threshold power.

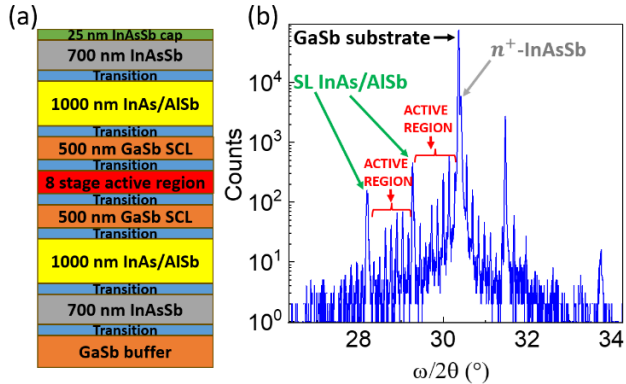

FIG. 3. (a) Detail layout overview of the ICL structure with hybrid claddings design (b) X-ray diffractogram of the ICL indicating lattice matching and strain compensated growth of high quality cladding layers

It is worth mentioning that the n$^+$-InAsSb is not transparent for wavelengths shorter than 4.3 µm. In order to match the band offsets in between the cladding layers, SCLs and the active region, transition regions composed of InAs/AlSb superlattices needed to be implemented. The structure was capped with 25 nm of InAsSb doped to $2 \cdot 10^{19}$ cm$^{-3}$ to reduce contact resistance. Figure 3 (b) shows a high-resolution X-ray diffraction (HR-XRD) scan of the grown ICL based on the hybrid cladding layer design. The GaSb substrate and outer InAsSb cladding peaks labeled in the figure show an excellent lattice matching with spacing of only 0.046º between them (mismatch of 720 ppm). Satellite peaks of the inner superlattice claddings are also highlighted. The 0$^{th}$ order SL reflex is coinciding with the GaSb substrate peak indicating that the inner strain of the superlattice was adequately compensated. Active region peaks show an excellent quality of the grown sample as they can be observed up to 20$^{th}$ order.

Broad area (BA) devices were fabricated for characterization in pulsed mode. For their processing, standard photolithography was carried out to define 100 µm wide ridges. In the wet etching process a mixture of $H_2O/H_3PO_4/H_2O_2/HOC(CO_2H)(CH_2CO_2H)_2$ was used to etch through the active region to the bottom SCL. Afterwards, a Ti/Pt/Au top contact was deposited and AuGe/Ni/Au layers were evaporated on a substrate as a back contact. The fabricated sample was cleaved into 2 mm long laser bars and mounted epilayer side up on a copper heat sink. Pulsed measurements were performed under a repetition rate of 1 kHz and the applied current pulse width was 500 ns to minimize Joule heating. Figure 4 (a) displays electro-optical characteristics of the BA ICL operated in pulsed mode under different temperatures (from 15 ºC to 55 ºC).

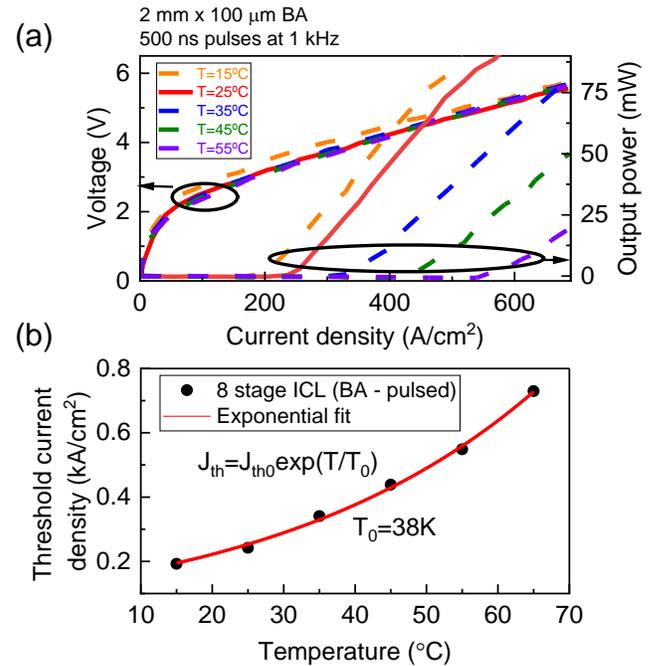

FIG. 4. (a) Pulsed light-current and light-voltage curves at different temperatures for a 2 mm long and 100 µm wide BA device. (b) Temperature dependence of the threshold current density and extracted characteristic temperature $T_0$ = 38 K.

Bandgap reduction induced by temperature shift causes a reduction of overall voltage drop per cascade, hence at the same current the voltage decreases at higher temperatures. At room temperature, the ICL exhibits a threshold current density of $J_{th}$ = 242 A/cm$^2$. The obtained differential resistance of 3 Ω is larger than the typical 1-2 Ω[22,29], which might be related to slight underdoping of the SCLs. Despite higher resistance, the threshold power density of $P_{th} = 840$ W/cm$^2$ is an excellent value for emission at 5.2 μm and comparable to the reported values at similar wavelengths[33]. The measured voltage efficiency of 55.4 % is comparable to reported values at similar wavelengths[29,32] but it could be improved by reducing the differential resistance. In the voltage range 0-2 V leakage current is present which could be a consequence of low sidewall resistance as no passivation was applied to the etched BA lasers. Additionally, there is a possibility of parasitic current channels due to misalignment of the energy states in the region of low electric field and voltage. For the displayed temperature range, the t efficiency drops from 144 to 66 mW/A. In Figure 4 (b) the temperature dependence of threshold current density is presented featuring a characteristic temperature of $T_0 = 38$ K, which compares well with reported values[19,28-29,34]. Figure 5 depicts the emission spectra of the ICL at three different temperatures. The temperature-induced wavelength shift is 4.4 nm/ºC.

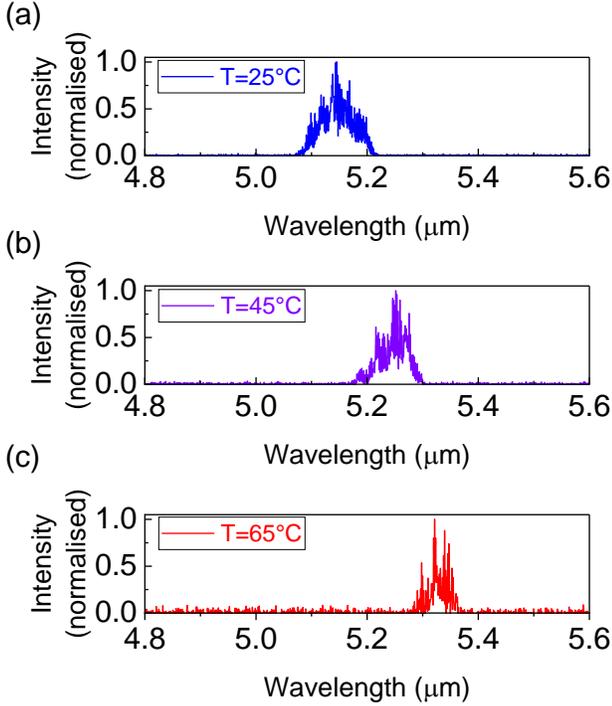

FIG. 5. Spectra at 25 ºC, 45 ºC and 65 ºC of a BA (2 mm x 100 μm) device. The central wavelength at room temperature is 5.16 μm and the measured temperature shift is 4.4 nm/ºC.

Ridge waveguides (RWG) were fabricated for analysis in cw operation. In contrast to the BA devices, these waveguides were patterned using e-beam lithography and fabricated using reactive ion etching. The materials used for both top and the backside contacts were identical to those employed for the BA devices. In addition, a layer of 11 μm gold was electroplated onto the top contact. Subsequently, the facets were coated with a high-reflective metal layer on the backside insulated by SiO$_2$ and Al$_2$O$_3$. 6.4 μm wide and 0.9 mm long Fabry-Pérot lasers were mounted epi-side up on copper heat sink. In Figure 6 (a) the electro-optical characteristics of RWG ICLs operated in cw regime are displayed at different temperatures. The maximum operational temperature was 41 ºC, approximately the same as highest reported in this wavelength range[15,29,34].

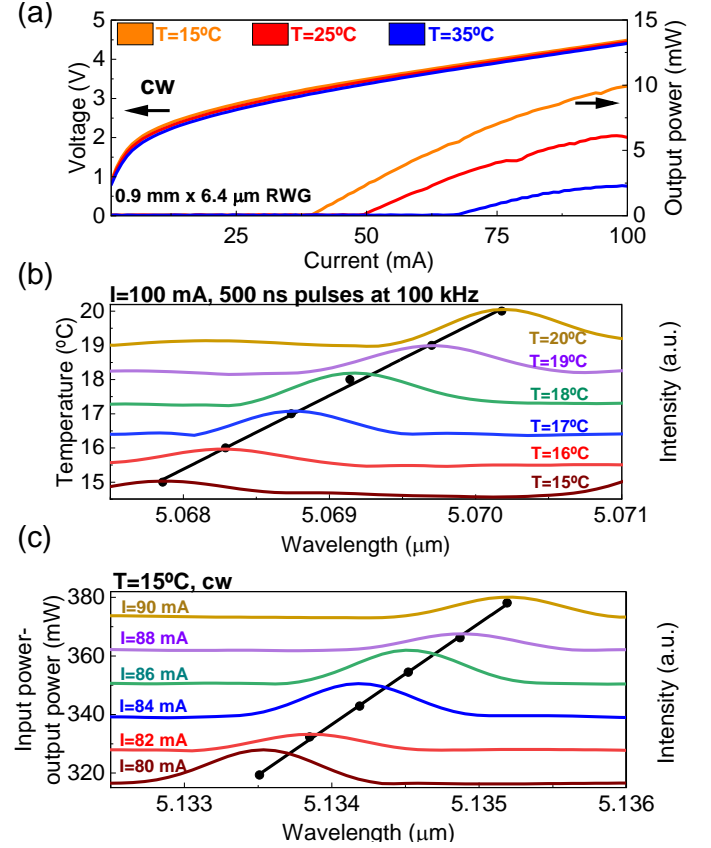

FIG. 6. (a) Continuous wave (cw) operation light-current and light-voltage curves at different temperatures for a 0.9 mm long and 6.4 μm wide ridge waveguide (RWG) device. (b) Temperature induced wavelength shift of a single Fabry-Pérot mode at pulsed operation of RWG laser. (c) Difference between input and output power of RWG laser for wavelength shift of a single Fabry-Pérot mode induced by sweeping of current at cw operation.

However, the threshold current at room temperature cw operation is 49.6 mA resulting in threshold current density of 861 A/cm$^2$ which is higher than previously reported values for ICLs at similar wavelengths[15,19,34]. This could be due to leakage currents caused by possible imperfections during the RWG fabrication process. The maximum measured output power of the ICL at 25 ºC was 6.1 mW resulting in the wall plug efficiency of 1.4 %, which is in the range of the reported values for ICLs of similar emission wavelengths[15,19,34]. Nevertheless, the presented results underline the promise of hybrid superlattice-plasmon-enhanced claddings as an alternative for commonly employed InAs/AlSb superlattice claddings. In Figure 6 (b) a temperature induced wavelength shift of a single Fabry-Pérot mode emitting in pulsed operation of a RWG ICL is depicted. In Figure 6 (c) the difference between input and output power of a RWG laser is shown at different wavelengths of a single Fabry-Pérot mode emitting in cw operation, at different currents and a fixed temperature of the heat sink. From the ratio of wavelength gradients of temperature and power displayed in Figures 6 (b) and (c), we obtain the thermal resistance of $R_{th} = 62$ K/W ($R_{th}A = 3.6$ K/(kW/cm$^2$)) which is significantly lower than the reported values[8,35] (100-150 K/W) for epi-side up RWG ICLs of similar cavity lengths. This is an indicator of enhanced heat dissipation.

In summary, we have demonstrated the lowest reported threshold current density for BA ICLs of wavelengths $\lambda \geq 5$ µm for a decent threshold power density. Optimization of the active region and an advanced hybrid cladding architecture was shown. Additionally, we have demonstrated a low thermal resistance of the RWG ICL wth hybrid plasmon-enhanced claddings in support to a good thermal management. Further improvement could be possible by alternative approaches in carrier-rebalancing as well as by additional optimization of optical confinement and loss management in terms of inner to outer cladding thickness ratio and doping.

We are grateful to European Union's Horizon 2020 research and innovation programme under the Marie Skłodowska-Curie grant agreement no 956548 (QUANTIMONY) for financial support of this work. We also thank S. Kuhn, S. Estevam and the production team of nanoplus Advanced Photonics Gerbrunn GmbH for sample processing and preparation.

The data that support the findings of this study are available from the corresponding author upon reasonable request.